\documentclass[12pt,preprint]{aastex}

\shorttitle{Testing Magnetic Star Formation Theory}
\shortauthors{Crutcher, Hakobian, \& Troland}

\begin{document}

\title{Testing Magnetic Star Formation Theory}

\author{Richard M. Crutcher, Nicholas Hakobian}
\affil{Astronomy Department, University of Illinois, Urbana, IL 61801}
\email{crutcher@illinois.edu, nhakobi2@astro.uiuc.edu}
\and
\author{Thomas H. Troland}
\affil{Physics and Astronomy Department, University of Kentucky,
    Lexington, KY 40506}
\email{troland@pa.uky.edu}

\begin{abstract}

Zeeman observations of molecular clouds yield the line-of-sight component $B_{LOS}$ of the magnetic vector {\bf B}, which makes it possible to test the two major extreme-case theories of what drives star formation -- ambipolar diffusion or turbulence.  However, only one of the three components of {\bf B} is measurable, so tests have been statistical rather than direct, and they have not been definitive. We report here observations of the Zeeman effect in the 18-cm lines of OH in the envelope regions surrounding four molecular cloud cores toward which detections of $B_{LOS}$ have been achieved in the same lines, and evaluate the {\em ratio} of mass to magnetic flux, $M/\Phi$, between the cloud core and envelope. This relative $M/\Phi$ measurement reduces uncertainties in previous studies, such as the angle between {\bf B} and the line of sight and the value of [OH/H]. Our result is that for all four clouds, the ratios ${\cal R}$ of the core to the envelope values of $M/\Phi$ are less than 1. Stated another way, the ratios ${\cal R^\prime}$ of the core to the total cloud  $M/\Phi$ are less than 1. The extreme case or idealized (no turbulence) ambipolar diffusion theory of core formation requires the ratio of the central to total $M/\Phi$ to be approximately equal to the inverse of the original subcritical $M/\Phi$, or ${\cal R^\prime} > 1$. The probability that all four of our clouds have ${\cal R^\prime} > 1$ is $3 \times 10^{-7}$; our results are therefore significantly in contradiction with the hypothesis that these four cores were formed by ambipolar diffuson. Highly super-Alfv\'{e}nic turbulent simulations yield a wide range of relative $M/\Phi$, but favor a ratio ${\cal R} < 1$, as we observe. Our experiment is limited to four clouds, and we can only directly test the predictions of the extreme-case ``idealized'' models of ambipolar-diffusion driven star formation that have a regular magnetic field morphology. Nonetheless, our experimental results are not consistent with the ``idealized'' strong field, ambipolar diffusion theory of star formation. Comparisons of our results with more realistic models and simulations that include both ambipolar diffusion and turbulence may help to refine our understanding of the relative importance of magnetic fields and turbulence in the star formation process. 

\end{abstract}


\section{Introduction}

Understanding star formation is a fundamental astrophysical problem. \citet{MO07} have comprehensively reviewed the field. For thirty years what has sometimes been called the ``standard model'' has been that magnetic fields control the formation and evolution of the molecular clouds from which stars form, including the formation of cores and their gravitational collapse to form protostars. However, in recent years doubts about the validity of this model have been raised by those who argue that turbulence controls the formation of clouds and cores, with cores either dissipating back into the general interstellar medium or collapsing and forming stars if they are self-gravitating when formed. In spite of decades of intense research, there is still not consensus on the role that magnetic fields play in the star formation process.  
 
Detailed theoretical work on the strong magnetic field ``standard model'' has been carried out by a number of groups: \citet{Setal99} and \citet{MC99} have reviewed and summarized the state of this theory. The fundamental principle is that clouds are formed with subcritical masses ($M < M_\Phi = \Phi/2\pi\sqrt{G}$). Here $\Phi$ is the magnetic flux and G is the gravitational constant, and the expression for $M_\Phi$ is from \citet{NN78}; other expressions for $M_\Phi$ differ slightly from this depending on cloud structure (e.g., \citealt{MS76}). The magnetic field is frozen only into the ionized gas and dust; neutral gas and dust contract gravitationally through the field and the ions, increasing mass (but not to first order flux) in the cloud cores. This process is known as ambipolar diffusion. When the core mass reaches and exceeds $M_\Phi$, the core becomes supercritical ($M > M_\Phi$), collapses, and forms stars. The magnetic flux remains behind in the envelope. Because the ambipolar diffusion time scale for the formation of supercritical cores is fairly long ($\geq 10^7$ yr), molecular clouds would have long lifetimes. The star formation efficiency (the ratio of mass in stars in a molecular cloud complex to the interstellar mass) is low (as observed) due to the slow rate of star formation and to the fact that much of the mass of the molecular cloud is left behind in the subcritical envelope as the core collapses.

The idea that star formation is primarily regulated by ambipolar diffusion was standard for many years.  However, doubts about the validity of this assumption were raised by the development of a weak-field, super-Alfv\'{e}nic model of dark clouds \citep{PN99}. The  new weak-field theory had molecular clouds being intermittent phenomena, with short ($\sim 10^6$ yr) lifetimes. In this theory clouds form at the intersection of turbulent supersonic flows in the interstellar medium. Generally, clouds do not become gravitationally bound and dissipate; those that are self-gravitating form stars in essentially a free-fall time \citep{E00}. Star formation occurs only in the small fraction of the molecular gas that is sufficiently dense to be self-gravitating. The star formation efficiency is low due to a small fraction of the mass of clouds becoming gravitationally bound. Magnetic fields are present in this theory, but they are too weak to be energetically dominant. The role of turbulence in the energetics of the interstellar medium has been a very active area.  \citet{Petal04} presented evidence in favor of the weak field, super Alfv\'{e}nic model. \citet{ES04} have written an excellent review of interstellar turbulence, and \citet{MK04} have extensively reviewed arguments that supersonic turbulence controls star formation.

Early work focused on extreme-case models, either strong magnetic field models that did not (at least directly) include the effects of turbulence, or strong turbulence models that neglected the effects of ambipolar diffusion. More recent theoretical work has introduced supersonic turbulence into numerical star formation models \citep[e.g.,][]{NL05, KB08, NL08}.  These more recent models apply to magnetically subcritical regions, and they include the effects of ambipolar diffusion.   The role of turbulence proves to be important in these models, significantly shortening, for example, the ambipolar diffusion timescale. In this paper we focus on a direct comparison of observational results with the strong field, ambipolar diffusion model without turbulence. Such models have a smooth morphology of the magnetic field which deforms into an hourglass shape with axis along the mean field direction, and make an analytic prediction that can be tested directly. The more complicated models that come from simulations that include both magnetic fields and turbulence can also be tested against the results of this paper, and we discuss how such tests might be carried out. 

Our approach to testing star formation theory has been to measure magnetic field strengths in molecular clouds in order to see whether they are weak or strong.  The crucial parameter is the ratio of the mass to the magnetic flux, $M/\Phi$, which is of course closely related to $M_\Phi$. If $M/\Phi$ is observed to be significantly supercritical, particularly at lower densities, the magnetic support theory is not viable. If it is observed to be subcritical at lower densities, magnetic fields would be too strong for the intermittent, turbulent theory to hold. The $M/\Phi$ parameter provides in principle a straightforward, direct test to discriminate between the two extreme theories of star formation.

The Zeeman effect provides the only known method to directly measure magnetic field strengths in dense gas. In this paper we first briefly review Zeeman results and argue that they have not yet provided an unambiguous discrimination between the two star formation theories. We then describe a new experiment to attempt to overcome the limitations of existing Zeeman results and to test directly a prediction of the ``standard model''. Finally, we describe the observations made for this experiment and the results, and discuss the implications of those results. 

\section{The Zeeman Effect}

\subsection{The Technique}
\label{zeetech}

If a spectral line forming region is permeated by a magnetic field {\bf B}, the line is split by the normal Zeeman effect into three separate frequencies, $\nu_0 - \nu_z$, $\nu_0$, and $\nu_0 + \nu_z$, where $\nu_z = B \times Z$, $B$ is the magnitude of {\bf B}, and $Z$ is the Zeeman coefficient in Hz/$\mu$G (note that $Z$ is often defined as two times our $Z$). Unfortunately, most molecules do not have large values of $Z$. In general, only those molecules with an unpaired outer electron will have a $Z$ of order the Bohr magneton, $M_B = eh/4\pi mc = 1.40$ Hz/$\mu$G. For the 1420 MHz H~I line, $Z = 1.4$ Hz/$\mu$G, so if $B = 1$ $\mu$G, the total splitting of the two circularly polarized components would be 2.8 Hz, or $6 \times 10^{-4}$ km s$^{-1}$. This very small magnitude of the Zeeman splitting makes Zeeman observations difficult and consuming of large amounts of telescope time. For most molecules, $Z$ will be of order the nuclear magneton, which is 1840 times smaller than $M_B$. Except for the very strong H$_2$O masers, all Zeeman work has involved the few species with large $Z$Õs: H I, OH, CN, CH, CCS, and SO, with Zeeman detections to date in only the first three species.

The observed data are Stokes parameter I and V spectra for each position. From these data we infer the column density N and the line-of-sight component $B_{LOS}$ of the magnetic vector {\bf B}. The I and V spectra have channel-to-channel noise that has a Gaussian probability density function (PDF) -- noise dominated by the stochastic receiver noise.  Hence, the random error in N is Gaussian, or very nearly so. For the usual case of Zeeman splitting much smaller than the line width ($\nu_z << \delta \nu$), only $B_{LOS}$ can be determined (e.g., \citealt{Cetal93}). The Stokes V spectrum actually consists of a possible scaled-down Stokes I signal (due to a possible gain difference between the two receivers) and a Zeeman signal that is proportional to the first derivative of the Stokes I spectrum. To infer $B_{LOS}$ we do a linear least-squares fit of $a \times I + b \times dI/dv$ to the observed Stokes V spectrum, where the Stokes I and V spectra are functions of radial velocity $v$, and $dI/dv$ is obtained by numerically differentiating the Stokes I spectrum. Both of the parameters $a$ and $b$ may be positive or negative. Hence, the position and shape of the Zeeman signal are set entirely by the Stokes I spectrum; only the amplitude (positive or negative) of the Zeeman signal is a free parameter. The sign of $b$ indicates whether the line-of-sight component of the magnetic vector {\bf B} points toward or away from the observer, and the magnitude of $b$ is proportional to the magnitude of $B_{LOS}$. The parameters $a$ and $b$ and the mean errors in each parameter come in the standard way from this straightforward linear least-squares fitting procedure. Hence, the error in $B_{LOS}$ that comes from the linear least-squares fit is expected to be Gaussian normal, or very nearly so. During the beginning years of Crutcher's observations of the Zeeman effect in OH lines, he confirmed by Monte Carlo tests that the linear least-squares fitting procedure gives correct results with very nearly Gaussian PDFs.

Since the magnitude of only one of the three components of the vector {\bf B} can be measured, in general it is necessary to apply statistical techniques in order to determine astrophysically meaningful parameters. If one observes a large sample of clouds distributed over the sky, the direction of the magnetic fields should be random with respect to the lines of sight from the observer. The usual approach is to assume that the total field strength $B$ is the same in an observed sample of clouds in order to infer $B$ from measurements of $B_{LOS}$. As discussed by \citet{HC05}, for this case both the median and mean values of $B_{LOS} = B/2$, so the mean and median of $B$ may be obtained for the observed sample of clouds. As \citet{HC05} noted, for PDFs for $B$ other than a delta function, the mean and median of $B_{LOS} \approx B/2$, so useful information about magnetic field strengths from Zeeman observations can be obtained even if the PDF of $B$ is unknown (as is of course the case).

Since the magnetic flux is just the magnetic field strength times the spatial area over which it is measured, $\Phi = B \times Area$. The mass within this area can be inferred from the hydrogen column density, $N_H = N(H I) + 2N(H_2)$; allowing for 10\% He, $M_{obs} = 1.4 m_H N_H \times Area$. Thus, $M/\Phi \propto N_H/B$. 

\subsection{Previous Zeeman Work}

Most of the earlier Zeeman detections in molecular clouds (e.g., \citealt{C99}) have been toward clouds associated with H II regions. Dark clouds offer the possibility of measuring the role of magnetic fields at an earlier stage of the star formation process, especially for the low-mass star formation case where the ``standard model'' may best apply. However, until recently there were only two dark cloud molecular Zeeman detections \citep{Getal89, Cetal93}. 
 
In order to improve our knowledge of magnetic field strengths in dark cloud cores, \citet{TC08} used the Arecibo telescope to carry out an extensive program to observe the Zeeman effect in the 1665 and 1667 MHz lines of OH. Thirty-three dark cloud core positions were observed. They achieved nine detections of $B_{LOS}$ and sensitive upper limits for the other positions. Also needed to compute $M/\Phi$ is an estimate for the column density of H$_2$. They obtained this estimate from the OH lines themselves. The Arecibo OH spectra yield N(OH). With OH/H = $4 \times 10^{-8}$ \citep{C79}, one can infer N(H$_2$). The \citet{TC08} Arecibo maps of OH emission around the core positions showed that OH does peak up on the CO and/or NH$_3$ cores, and that OH samples densities up to around n(H$_2$) $\sim 2 \times 10^4$ cm$^{-3}$. 

Although the \citet{TC08} analysis of their data gave the mean value of $B$ for their dark-cloud sample, it provided no information about the possible variation of $B$ from core to core. They inferred $\overline{\lambda}$, the ratio of the observed mean $M/\Phi$ to the critical value. The data were consistent with the prediction of the strong magnetic field theory -- a slightly supercritical $M/\Phi$ if the core morphology is that of a disk, as predicted by the strong-field theory.

The Arecibo Zeeman observations had the potential to eliminate one of the two extreme-case theories for the star formation process. If $M/\Phi$ had been found to be unambiguously highly supercritical, the ambipolar diffusion driven theory would have been eliminated. If $M/\Phi$ had been found to be unambiguously subcritical, the turbulence driven theory would have been eliminated. The statistical result was that the mean $M/\Phi$ in molecular clouds was observed to be approximately critical, i.e., $\overline\lambda \approx 2$. Because of the many non-detections, and since the OH/H$_2$ ratio is uncertain by a factor $\sim 2$, a mean $M/\Phi$ ranging from critical to supercritical by $\sim 4$ was consistent with the data. Of course, there could be a real variation with an unknown range in $M/\Phi$ from cloud to cloud. Hence, although the observations have shown that magnetic fields are sufficiently strong that they cannot be ignored, the very hard-won observational results cannot rule out either extreme-case theory of star formation. 

\section{This Experiment}

The goal of the experiment described in this paper is to perform a definitive test of the extreme-case (no turbulence) ambipolar diffusion theory of star formation that circumvents the ambiguities inherent in previous Zeeman observational tests. In addition, the experiment will supply valuable observational data against which more complicated models (involving both ambipolar diffusion and turbulence) may be tested. This definitive test is measurement of the change in $M/\Phi$ between the envelope and the core of a cloud. The magnetic support/ambipolar diffusion theory makes a specific prediction that can be tested. It requires that $M/\Phi$ of the original cloud be subcritical, with ambipolar diffusion accumulating mass but not flux in the cloud center, building up a higher density core. Eventually, the core becomes supercritical and starts a collapse which is slower than free fall due to the magnetic pressure. During this supercritical collapse phase, magnetic flux is dragged inward, so the rate of increase of $M/\Phi$ is slowed. The result is a prediction that the ratio of the core to the original cloud $M/\Phi > 1$, by approximately the inverse of the amount by which the original cloud was subcritical. 

Is the prediction of the ambipolar diffusion theory sufficiently different from the expectations of the turbulence-driven theory? The answer is clearly yes, for the turbulence theory generally predicts the opposite behavior of the mass-to-flux ratio between envelope and core. Simulations of the formation of cores by turbulence \citep{EVS05} made the usual assumptions for this theory of initially uniform density and magnetic field, driven turbulence, and ideal MHD (strict flux freezing, i.e., no ambipolar diffusion).  They found that $M/\Phi$ usually decreased with increasing density, the opposite of the ambipolar diffusion result. A change in $M/\Phi$ with ideal MHD seems impossible, but it occurs due to the way $M/\Phi$ is measured. No change in $M/\Phi$ occurs when an entire flux tube is considered. But converging flows perpendicular to {\bf B} increase $B$ and $M$ within a fixed volume, while flows parallel to {\bf B} can increase or decrease $M$ within that fixed volume. Hence, when measuring $M/\Phi$ within the volume chosen to define a cloud or core, $M/\Phi$ can be either larger or smaller than the original value. \citet{EVS05} noted that for mass and magnetic flux conservation, a clump or core within a cloud has a larger density than the mean, but a mass smaller than its parent cloud, so a core would generally have a smaller $M/\Phi$ than the parent cloud (since the magnetic flux would be unchanged by core formation but the core mass would be only a fraction of the mass within the flux tube). Hence, this turbulent simulation result may be general, although of course it would require more studies to verify that conclusion.

Strong magnetic field cases can obtain a similar result. If the mass within a flux tube fragments into several cores on a shorter time scale than the ambipolar diffusion one, each core will have only a fraction of the mass in the flux tube, and the $M/\Phi$ of these cores would have decreased with respect to that of the original material \citep{M91}. Although the physics driving this fragmentation is different from that considered by the turbulence theory, the fundamental reason for the decrease in $M/\Phi$ is the same, and it is not related to ambipolar diffusion. Here we address only testing the ambipolar diffusion driven formation of dense cores, for which $M/\Phi$ must increase.

A quantitative test of a theoretical prediction requires the design of an experiment to test specific numerical models. We designed our experiment to test the ``idealized'' models that have been published by the ambipolar diffusion theorists. We call these models idealized because they do not follow the evolution in a consistent way through the formation of clouds from a diffuse interstellar medium to the formation of cores in those clouds driven by ambipolar diffusion. Instead, the models generally start with an isolated, uniform, spherical cloud that is allowed to relax to an equilibrium state, which then evolves due to ambipolar diffusion. The non-thermal motions and irregular structures of observed clouds, which themselves are embedded in a complex interstellar medium, are not directly considered in these idealized models. So the models are two-dimensional (with azimuthal symmetry about the magnetic field direction) with regular (non-twisted) magnetic fields. To design the experiment, we looked at three such idealized ambipolar diffusion models: (1) a dimensionless parameter model \citep{CM94} for which they listed specific physical parameters for comparison with actual molecular clouds; this model with their listed physical parameters had an unevolved cloud radius of 4.3 pc; (2) one specifically computed for L1544 \citep{CB00} with an unevolved cloud radius of about 2.5 pc; and (3) a specific model for B1 \citep{Cetal94} with an unevolved cloud radius of 2.9 pc. These radii would become the radii of the ``envelopes'' surrounding the cores formed by ambipolar diffusion, since the region of the cloud outside the cores would be ``held in place'' (an ambipolar diffusion theorist phrase) by their subcritical magnetic fields. Moreover, the radii of the cores in these models were all $\sim 0.1$ pc. So we need to sample the core with a filled beam of radius $\sim 0.1$ pc and the envelope with a beam radius less than $\sim 2$ pc that excluded the core. A second aspect of the idealized  ambipolar diffusion models was that the magnetic fields were smooth and regular within their unevolved cloud radii, although with an ``hourglass'' morphology strongest in the core. Although this regularity of the field was by construction in the models, the requirement that $M/\Phi$ in the clouds be subcritical meant that the magnetic fields must be strong, so that even if turbulence had been included in the models, magnetic energy would likely dominate turbulent energy and field lines would in fact be quite regular. Hence, these two inputs from the actual ambipolar diffusion models guided our design of the experiment -- we needed to sample cores and envelopes on the relevant spatial scales and we could (to first order) assume that the magnetic fields in the clouds were not significantly twisted but mainly ordered. 

We chose four clouds for our experiment -- two in the Taurus molecular cloud complex (distance $\sim 150$ pc) and two in the Perseus molecular cloud complex (distance $\sim 300$ pc). A 0.1 pc core radius would be  $\sim 1^\prime (2^\prime)$ and a 2 pc cloud (or envelope) radius would be $\sim 20^\prime (40^\prime)$ at the 300 (150) pc distance. At the OH line frequency, the primary-beam radius of the Arecibo telescope is $\sim 1.5^\prime$, so this is well matched for measurement of core properties at the densities sampled by OH. The Green Bank telescope (GBT) beam radius is $\sim 3.9^\prime$; by pointing the GBT at positions $6^\prime$ from the Arecibo pointing position, the GBT beams would exclude the molecular core and sample (at half-power response) the radius range $2.1^\prime$ to $9.9^\prime$, or $\sim 0.2 (0.1)$ pc to $\sim 0.9 (0.45)$ pc at the 300 (150) pc distance. Figure 1 displays these beam sizes and positions. This choice of beams was set by the specifications of the ambipolar diffusion models cited above. The sampling of the envelope had to be sufficiently far from the core to obtain a significantly different result from the core result, but {\em well} within the outer boundary of the unevolved clouds; the GBT beamsize was ideally suited to these objectives. We could then ``synthesize'' a toroidal or ring beam to sample the envelopes by appropriately combining the observations from the four GBT beams in order to produce exactly the envelope sampling called for by the ambipolar diffusion models. 

The Arecibo OH survey of dark clouds produced detections of $B_{LOS}$ in 9 of 33 cores observed; earlier, \citet{Getal89} obtained a detection with Arecibo toward the B1 core. The Arecibo observations have therefore selected 10 cores for which there are measurements (not just limits) for $N_{OH}/B_{LOS}$ ($\propto M/\Phi$) with a $3^\prime$ beam, well matched to the core diameters of $\sim 2^\prime$. Additional observations that are needed are measurements of  $N_{OH}/B_{LOS}$ in the envelopes of these cores to test the prediction that ambipolar diffusion increases the mass but not (at least very much initially) the field in cloud cores. Zeeman observations in molecular envelopes have generally not been attempted before, since line strengths are weaker and Zeeman detections are difficult to obtain. 

In order to measure $M/\Phi$ in the envelopes surrounding the Arecibo cores, for this experiment we made a four-point Zeeman map with the GBT at positions $6^\prime$ from each of four cores with detected $B_{LOS}$ (see figure 1), dividing the observing time about equally between the four positions. Combining the four envelope results for each cloud can then give results that would be obtained with the required toroidal or ring telescope beam; of course, sensitivity around the ring is not uniform due to four telescope beams that make up the ring. This sampling of the envelope regions is exactly what we require to measure $M/\Phi$ in the envelopes, excluding the cores. Therefore, we have $B_{LOS}$ and $N_{OH}$ separately for the core and envelope material. The ratios $[N_{OH}/B_{LOS}]_{core}$ and $[N_{OH}/B_{LOS}]_{envelope}$ will then be available for the clouds. The strong field/ambipolar diffusion theory for core formation requires $[M/\Phi]_{core}/[M/\Phi]_{envelope} =  [N_{OH}/B_{LOS}]_{core}/[N_{OH}/B_{LOS}]_{envelope} > 1$ (that is, that $M/\Phi$ increases from envelope to core). So these observations could result in the ambipolar diffusion theory being proved wrong if the observations find this core/envelope $M/\Phi$ ratio to be less than one. It could also result in the driven turbulence, ideal MHD simulations being proved wrong, if this experiment finds this core/envelope $M/\Phi$ ratio to be greater than 1. It is of course impossible to prove that a theory is correct, but this experiment has the potential to rule out one of the two competing extreme-case theories of star formation. Such a result would have a profound effect on further theoretical work on star formation, and would be a major advance in understanding the star formation process.

We define
\begin{equation}
{\cal R} \equiv \frac{M_{core}/\Phi_{core}}{M_{envelope}/\Phi_{envelope}}.  \label{eq: R}
\end{equation}
The mass (the OH lines are optically thin) is given by 
\begin{equation}
M \propto I \; \Delta V\;  A   \label{eq: mass}
\end{equation}
and the magnetic flux by
\begin{equation}
 \Phi \propto (B_{LOS}/cos \theta)\; A.  \label{eq: flux}
 \end{equation}
 
 Here $I$ is the peak intensity of the spectral line, $\Delta V$ is the FWHM line width, $A$ is the area of sky sampled, $A = \pi r_c^2$ for the core and $A = 4 \pi r_e^2$ for the envelope, with $r_c$ and $r_e$ being the radii of the telescope beams used respectively for the core and envelope observations, $B_{LOS}$ is the magnitude of the line-of-sight component of the magnetic field, and $\theta$ is the angle between the line of sight and {\bf B}. Since $A$ is the same for measuring $M$ and $\Phi$ for the core, it divides out of the numerator of ${\cal R}$; similarly in the denominator for the envelope. The factor $cos\theta$ also divides out.

Theoretical calculations for ambipolar diffusion generally do not predict ${\cal R}$ as we have defined it, but rather $M/\Phi$ in the central flux tube as a function of the central density, which increases with time (e.g., \citealt{CM94}, figure 2e). For a disk morphology with the field along the line of sight, the central flux tube would include contributions from the foreground and background parts of the envelope, but these would be small and the central result would be dominated by the core $M/\Phi$. In any case, our measurement of $(M/\Phi)_{core}$ corresponds to what the models give. Hence, for a quantitative comparison with those predictions, we would need $M/\Phi$ through the center of the initial cloud before ambipolar diffusion had created the core. Before ambipolar diffusion has acted, the initial central $M/\Phi$ is equal to the $M/\Phi$ of the initial cloud as a whole. We can infer that original central $M/\Phi$ from our observations. The total mass and flux of the cloud do not change as the cloud evolves due to ambipolar diffusion, so the present $(M/\Phi)_{core + envelope}$, obtained by adding together the Arecibo result for the core and the four results for the envelope, gives the initial central mass-to-flux ratio $(M/\Phi)_{c0}$ used as the starting point of the ambipolar diffusion calculations. We therefore define
\begin{equation}
{\cal R^\prime} \equiv \frac{M_{core}/\Phi_{core}}{M_{core + envelope}/\Phi_{core + envelope}}.  \label{eq: R'}
\end{equation}
${\cal R^\prime} \approx (M/\Phi)_{c}/(M/\Phi)_{c0}$, the quantity predicted by ambipolar diffusion models (e.g., \citealt{CM94}). The expressions in equations~\ref{eq: mass} and~\ref{eq: flux} for $M$ and $\Phi$ are used in equation~\ref{eq: R'}, but now the areas $A$ do not divide out in the denominator and explicit values must be used. The factor $cos\theta$ still divides out, however.

Measuring relative values of $M/\Phi$ ($\cal{R}$ or ${\cal R^\prime}$) will eliminate or at least greatly reduce the uncertainties inherent in absolute measurements of $M/\Phi$. We seek only a change in $M/\Phi$ from envelope to core, not the absolute values themselves. This avoids all of the geometrical correction problems in going from $B_{LOS}$ to the total $B$ and from $N_{obs}$ to $N_B$ (the column density along {\bf B}, which is what is required to properly compute $M/\Phi$). We do not need to know  the angle $cos \theta$ between the line of sight and {\bf B}, because we will be making a relative measurement, so $cos \theta$ in equation~\ref{eq: flux} will divide out. Similarly, no geometrical correction for the measured column densities will be necessary. Also, by observing $B_{LOS}$ in the core and envelope using the same tracer, in this case OH, the ratio OH/H that would be needed to convert measured OH column density to total (H$_2$) column density in order to find the absolute value of $M/\Phi$ is not needed. $\cal {R}$ and ${\cal R^\prime}$ are inferred from the directly measured quantities $I$, $\Delta V$, $B_{LOS}$, and the radii of the telescope beams used for the core and envelope observations.

\section{Observations}

\subsection{Arecibo Observations\label{Arecibo}}

The Arecibo\footnote[1]{The Arecibo Observatory is part of the National Astronomy and Ionosphere Center, which is operated by Cornell University under a cooperative agreement with the National Science Foundation.} observations of cores used here are not new observations, but have been reported already: B1 \citep{Getal89}, L1544 \citep{CT00}, B217-2 and L1448CO \citep{TC08}. Details of these observations are described in these papers. These core positions were selected from catalogs of molecular peaks observed in CO, NH$_3$, and other tracers. The exact positions for which OH Zeeman observations were carried out were refined by small Arecibo OH maps to determine the peak OH line strength positions. The B1 observations were carried out with the old Arecibo line feed, while the others used the new Gregorian feed. The spectral resolution of the B1 observations was lower due to the older spectrometer in use at that time, but the line is well resolved so this does not affect the line strength and the Zeeman measurement. The Gregorian feed data were originally not calibrated exactly correctly. This did not affect the determination of $B_{LOS}$, which depends on the ratio of the Stokes I and V spectra that would have been affected precisely the same way. But for this study, we re-calibrated the spectra using the best known value of the L-band feed noise diode. This resulted in about a 7\% reduction in the line strengths from those published previously.

The characteristics of the Arecibo telescope beam have been carefully studied by \citet{Hetal01a}. This study was of the Gregorian feed, but the essential characteristics of the line feed for the purposes of this paper are not significantly different. One important parameter is the main beam diameter at the half-power point, $2.9^\prime$. This is the size of the core region sampled by the Arecibo observations. Second, there is the main beam efficiency, which is only 0.48. This means that slightly more than half the area of sky to which the Arecibo observations were sensitive lies outside the $2.9^\prime$ diameter of the main beam, mostly in sidelobes. \citet{Hetal01a} measured the telescope response out to the first sidelobe, which lies $\sim 5^\prime$ from the beam center, close to the positions where the GBT envelope observations were centered. This first sidelobe has 0.14 of the spatial response of the telescope, so in addition to the response of the Arecibo telescope to the cores, there is a response to the envelope emission equal to 0.14/0.48 = 29\% of the core response. This first sidelobe is a ``ring'' beam similar to the pseudo-ring beam formed by the four telescope beams used for the envelope observations (see figure 1). Hence, for two reasons, the Arecibo results for $(M/\Phi)_{core}$ are ``contaminated'' by input from the envelope region. First is the fact that the Arecibo beam must pass through the near and far regions of the envelope along the line of sight. This is okay, however, for the theoretical models predict $M/\Phi$ for the central flux tube that would include the near and far envelope. The second is the ``ring'' beam produced by the first sidelobe. The effect of this would be to make $(M/\Phi)_{core}$ look more like $(M/\Phi)_{envelope}$ than its ``real'' value, i.e., to make ${\cal R}$ and ${\cal R^\prime}$ closer to 1 than either would be if the core only could be sampled. Without much more detailed information, it is impossible to deconvolve the envelope contribution out of the measured $(M/\Phi)_{core}$. This means that a difference between a measured ${\cal R}$ or ${\cal R^\prime}$ and 1 is more statistically significant than the significance determined solely from the measurement uncertainties.

\subsection{Green Bank Telescope Observations}
 
The observations of the envelope positions were carried out with the L-band receiver of the NRAO\footnote[2]{The National Radio Astronomy Observatory is a facility of the National Science Foundation operated under cooperative agreement by Associated Universities, Inc.} 100-m Robert C. Byrd Green Bank Telescope (GBT) between Oct 2007 and Apr 2008. The main beam diameter of the GBT at 1666 MHz is $7.8^\prime$ and the main beam efficiency is 0.81; these quantities come from our own ``spider'' scan polarization calibration observations (see below). A negligibly small fraction of the GBT response comes from the core region of each cloud, due to the small sidelobe intensity, the fact that the cores lie near the first null of the beam response, and the fact that the core fills only a small fraction of the $360^\circ$ surrounding the GBT beam center. Observing time at each of the 16 positions ($6^\prime$ north, south, east, and west of each of four Arecibo cores, see figure 1) was approximately equal, averaging about 9.5 hours of actual on-source Zeeman integrations per position. The NRAO Spectral Processor was used as the backend correlation spectrometer and polarimeter. Simultaneous observations were made in both horizontal and vertical linear polarization of the 1665 and 1667 MHz OH lines. The total spectrometer bandwidth for each line in each polarization was  156.25 kHz or about 28 km~s$^{-1}$, which yielded a spectral channel width of about 0.055 km~s$^{-1}$. Bandpass calibration was carried out by frequency switching between -15/64 and +15/64 of the total bandwidth; after appropriately combining the two halves of the band, this resulted in a final velocity coverage for each spectrum of about 13 km~s$^{-1}$. The initial data processing made use of IDL software written by C. Heiles and T. Robishaw, which provided calibrated Stokes I, V, Q, and U spectra. Linear baselines were removed from the Stokes I and V spectra. The polarization calibrations used ``spider'' scans of the continuum source 3C286 over several hours surrounding its transit. \citet{Hetal01b} provide details of the polarization calibration procedure. In addition to the molecular cloud envelope observations, we briefly observed the polarized OH maser sources W49 and W3(OH) in order to verify the observing setup and data processing procedures. In addition, we observed the previously well-studied OH Zeeman absorption line source NGC2024 (Orion B, W12) in order to verify that the procedures produced the same $B_{LOS}$, including the sign or direction of the magnetic field, as had been obtained previously \citep{CK83, HS86, Cetal99, Betal01}. (Note that the field direction reported in the \citet{CK83} paper was wrong; the correct direction is given in the latter three papers.) The GBT Zeeman results for NGC2024 agree with the previous results.

Although the Heiles-Robishaw software produces a Zeeman fit to the spectra, it does so over the full bandwidth. The Stokes I spectra at the molecular core positions are dominated by a single strong component, but at the envelope positions this component is weaker and additional components are present that are sometimes comparable in strength with the core component. We therefore used our software, the same code used previously to fit the Arecibo survey data \citep{TC08}, to fit for $B_{LOS}$ only over those spectral channels in which the strong core line component dominates; these are the results reported here. Details of the procedure are given by \citet{TC08} and briefly described in \S\ref{zeetech} above. 

There are several possible methods to combine the results for the four envelope positions surrounding each core in order to produce a measurement of $B_{LOS}$ for each envelope. These include (1) taking the mean of the four envelope results for $B_{LOS}$, or (2) first averaging the Stokes I and V spectra for the four envelope positions before fitting for $B_{LOS}$. The second method has the disadvantage that the radial velocities and line widths are not the same at each envelope position, so the averaged spectra would be broader than any of the individual spectra, which would slightly reduce the sensitivity to $B_{LOS}$. We feel that instead of either of these two, the best analysis technique is to fit all eight spectra (the 1667 and 1665 MHz spectra for the four envelope positions for each cloud) {\it simultaneously} in order to obtain the best-fit single value for $B_{LOS}$ and its uncertainty from the entire envelope data set. This preserves the information content of each of the eight spectral lines, so no broadening takes place that would reduce the sensitivity. It also imposes the constraint that a single value of $B_{LOS}$ describes all of the data, which is the desired result for a synthesized toroidal beam measurement of $B_{LOS}$ in each envelope.

\section{Results}

Results for the core and envelope observations are shown in table 1 and in figures 2-5. Table 1 gives the numerical values for $I_{OH} = T_A(1665) + T_A(1667)$, $\Delta V$, and $B_{LOS}$. The values listed for the envelopes are the results from the simultaneous fit for a single $B_{LOS}$ and its mean error over both OH lines and all four envelope positions. Plotted in each panel of figures 2-5 is the observed antenna temperature $T_A$ of the 1667 MHz line at each of the observed positions, in order to show the relative line strengths and widths at the core and the envelope positions. The 1665 MHz line strengths are typically about 60\% of the 1667 MHz line strengths and show the same relative strengths from position to position. In the upper left of each panel is $B_{LOS}$ and its $1\sigma$ uncertainty for that position. Although the line strength does not decrease uniformly in all directions from the cores, and in most cases there is OH emission at velocities slightly away from the velocity of the main core component, the line strength in the envelope is typically about 50\% that of the core. Moreover, in all cases, the line width at the core position is only about 75\% that at the envelope positions. A decrease in line width from envelope to core is a standard feature of molecular clouds, seen even more strongly in molecular tracers that sample higher densities than does OH. So both the increase in line strength and the decrease in line width from the GBT to the Arecibo spectra argue that the Arecibo OH observations do sample molecular cores, and that the GBT observations are dominated by non-core emission. 

$B_{LOS}$ at L1544west is twice $B_{LOS}$ at L1544core in spite of the OH column density ratio being 0.6; this is the only case of a stronger field at an envelope position than in the core. Toward B1 there appears to be a north-south ridge of $B_{LOS}$ with much weaker $B_{LOS}$ to the east and west of the core. At all of the remaining 13 envelope positions the signal/noise ratio is low, although at all 16 positions the sensitivity to $B_{LOS}$ is sufficient to provide significant results for ${\cal R}$ and ${\cal R^\prime}$. A hypothetical measured envelope $B_{LOS} = 0$ is a perfectly acceptable experimental result for computing ${\cal R}$ and ${\cal R^\prime}$; the only requirement is that the sensitivity to $B_{LOS}$ be sufficient to yield statistically meaningful results for ${\cal R}$ and ${\cal R^\prime}$.

Table 2 shows ${\cal R}$ and ${\cal R^\prime}$ and the $1\sigma$ uncertainties for each cloud. The uncertainties in ${\cal R}$ and ${\cal R^\prime}$ depend on the uncertainties for the $B_{LOS}$ given in table 1 and the uncertainties in $I_{OH}$ and $\Delta V$. The uncertainty in the $\Delta V$ is about 0.02 km s$^{-1}$ in each case. The nominal uncertainty in $I_{OH}$ from the channel-to-channel noise in the spectra is only about 0.01 K. The uncertainties in both $I_{OH}$ and $\Delta V$ are too small to contribute to the uncertainties in ${\cal R}$ and ${\cal R^\prime}$. However, the absolute calibration of the line strengths is uncertain by about 10\%. Although this systematic uncertainty would divide out in ${\cal R}$ and ${\cal R^\prime}$ if the same telescope had been used for all observations, it may be that there is a calibration difference between the Arecibo telescope and the GBT. We have therefore used an uncertainty of 10\% in each of the core and envelope values for $I_{OH}$. Even so, this uncertainty is insignificant in the error budgets for ${\cal R}$ and ${\cal R^\prime}$. Those error budgets are dominated by the uncertainty in the $B_{LOS}$ of each envelope, since the signal/noise ratio for those quantities is by far the lowest of any of the four measured quantities that go into ${\cal R}$ and ${\cal R^\prime}$. We assume that the PDFs of these four measured quantities are Gaussian normal distributed (see \S \ref{zeetech}). 

We obtained the results for ${\cal R}$ and ${\cal R^\prime}$ and their $1\sigma$ uncertainties with two different methods. First, we compute the uncertainties in ${\cal R}$ and ${\cal R^\prime}$ by normal error propagation. This procedure is justified by the fact that the uncertainties in $B_{LOS}$ are Gaussian normal (or very nearly so), and the possible systematic uncertainty in the N contribute insignificantly in comparison with the uncertainties in the $B_{LOS}$. However, due to the low signal-to-noise ratio in the $B_{LOS}$ (especially in the envelope field strengths), the assumption of Gaussian normal errors in ${\cal R}$ and ${\cal R^\prime}$ may not be strictly correct. We have therefore computed the PDF for ${\cal R}$ and ${\cal R^\prime}$ for each of the four clouds by Monte Carlo simulations; for each Monte Carlo simulation, $10^6$ trials were used. The results for ${\cal R}$ are shown in figure 6; results for ${\cal R^\prime}$ are extremely similar. Note that ${\cal R}$ and ${\cal R^\prime}$ may be positive or negative. Over plotted on the Monte Carlo PDFs are the Gaussian normal error curves with the parameters given by the error propagation method. In all cases, the means and standard deviations from the two methods are essentially identical. However, the Monte Carlo PDF generally has a stronger tail at the high values of ${\cal R}$ and ${\cal R^\prime}$, which slightly increases the probability that ${\cal R}$ (and ${\cal R^\prime}$) are greater than one. These probabilities are essentially identical for ${\cal R}$ and ${\cal R^\prime}$, so we list only one probability for each cloud in table 2. Each of the four clouds is unlikely to have ${\cal R}$ and ${\cal R^\prime} >  1$, with probabilities of being greater than one varying from less than 1\% to about 10\% (table 2). The probability that all four of our clouds have ${\cal R^\prime} > 1$ is just the product of the probabilities in table 2, or $3 \times 10^{-7}$, a highly significant result. Our experiment is therefore in contradiction with the hypothesis that these four cores were formed by ambipolar diffuson.  

There are several possible biases that could affect ${\cal R}$ and ${\cal R^\prime}$. First, as discussed above, the Arecibo telescope beam pattern biases the observed ${\cal R}$ and ${\cal R^\prime}$ to be closer to 1 than the ``real'' values. Since the observed ${\cal R}$ and ${\cal R^\prime} < 1$, these observed values are biased to be higher than the actual values.    In addition, there is a factor that may systematically raise or lower ${\cal R}$ and ${\cal R^\prime}$ -- the possible curvature of the magnetic field lines. Field lines will be drawn into an ``hourglass'' morphology as a core forms. Because the four cores out of 34 that we have observed in this experiment are among the few with strong $B_{LOS}$, it is likely that {\bf B} points approximately along the line of sight for these four cores. The field line through the cloud center would then point along the line of sight, so $B_{LOS}$ would equal the total $B$. However, other field lines passing through a core would be curved away from the line of sight in the near and far side of the core, so $B_{LOS}$ would be less than the total $B$. In the envelope a similar effect would hold, but the curving of magnetic field lines in the ``hourglass'' would be less extreme in this lower density outer region. See \citet{GS93} figure 5b for an example of the morphology of magnetic field lines; for a typical distance of 150 pc, the radius of the Arecibo beam would be $2 \times 10^{17}$ cm. Hence, one might generally expect $B_{LOS}$ to underestimate $B$ by a smaller factor in the envelope than in the core of a cloud, which would bias ${\cal R}$ and ${\cal R^\prime}$ to be too high -- the same sense as the Arecibo sidelobe bias. However, exactly which direction the field curvature bias would go and by how much would depend on the detailed structure of the magnetic field morphology. But in any case, the bias must be insignificant. Even if the mean field directions over the GBT and Arecibo beams differed by $30^\circ$ due to an hourglass morphology, a very unrealistically large value, the change in the measured $B_{LOS}$ would be by the factor $cos 30^\circ = 0.87$, and ${\cal R}$ and ${\cal R^\prime}$ would be biased high by the factor 1/0.87. The resulting ${\cal R}$ and ${\cal R^\prime}$ would be changed by only about $1\sigma$ by such an unrealistically large net angle difference. (Of course, if the magnetic fields are not regular, this angle could be much larger, but for our comparison with the ``idealized'' ambipolar diffusion model this is not relevant.) Finally, since we assume that $N(OH) \propto$ the mass $M$, if [OH/H] varies between the envelope and core, ${\cal R}$ and ${\cal R^\prime}$ would be affected. It seems unlikely that there is a significant systematic variation in [OH/H] between the envelopes and cores that we have observed. There is only a 50\% difference in the column density of OH between core and envelope, so the physical regions being sampled are not that different. Moreover, \citet{C79} found no evidence for a variation in [OH/H] up to $n(H_2) \sim 2 \times 10^4$ cm$^{-3}$, which includes the range our observations sample. So our measured ${\cal R}$ and ${\cal R^\prime}$ can be slightly biased, probably overall in the direction of being closer to 1 than the ``real'' values, but not by sufficiently large amounts to change the significant result of this experiment.

\section{Discussion}

We have found that $M/\Phi$ decreases significantly from envelope to core, or from the initial central value to the present evolved central value, for the clouds we have studied. How does this observational result compare with the prediction of the strong field, ambipolar diffusion driven theory and the weak field, turbulence driven theory?

We first consider the ambipolar diffusion theory. We compare our results with what we called above idealized models, which include only gravity, regular magnetic fields, and thermal pressure. The non-thermal motions and irregular structures of observed clouds, which themselves are embedded in a complex interstellar medium, are not directly considered in these idealized models. Perhaps for this reason, the ambipolar diffusion models are not compatible with the structure we observe in both column density and magnetic field strength. Figures 2-5 show clearly that the column densities vary considerably around the core positions of our clouds, rather than the uniform result that would be expected from a idealized model with {\bf B} oriented closely along the line of sight. Also, the observed $B_{LOS}$ in two envelopes clearly show structure, with $B_{LOS}$ at the west envelope position of L1544 being about twice as strong as toward the core, and much stronger than toward the other three envelope positions. B1 shows a north-south ``ridge'' of $B_{LOS}$, with weaker $B_{LOS}$ toward the east and west envelope positions. These observations cannot be reproduced by an ambipolar diffusion idealized model. 

The strong magnetic field/ambipolar diffusion theory {\em requires} that $M/\Phi$ increase in the core as evolution proceeds; after all, this increase {\em is} ambipolar diffusion, the heart of this theory. Hence, this theory predicts ${\cal R^\prime} > 1$. The amount by which ${\cal R^\prime}$ exceeds 1 would be dependent on the specific parameters for a model cloud -- mainly the assumed initial $M/\Phi$. For example, \citet{CM94} discussed a model cloud broadly consistent with the parameters of the clouds we observe. Although their model is in terms of dimensionless parameters, they state that the original unevolved cloud is consistent with a temperature of 10 K, a central density of  $2.6 \times 10^3$ cm$^{-3}$, a central magnetic field of 35 $\mu$G, a radius of 4.3 pc, and a total mass of 98 $M_\odot$. At the time when the core becomes magnetically critical, the central density has increased by a factor $\sim 37$ while the central $B$ has increased by less than 1.7. The initial $M/\Phi$ is subcritical with the central $M/\Phi = 0.256$ of critical; for this model, the predicted ${\cal R^\prime} \approx 4$. However, our results for all four observed clouds is ${\cal R^\prime} << 4$ with high degrees of significance.

There have been ambipolar diffusion models calculated specifically for two of the clouds in our sample, for comparison with observational data available at that time. \citet{Cetal94} discussed a model specifically for B1; it had a core mass of 13 $M_\odot$, an envelope mass of 600 $M_\odot$, an envelope radius of 2.9 pc, an initial central $B = 43$ $\mu$G, and an initial central $M/\Phi = 0.42$ of critical. The model assumed that the cloud was a disk whose minor axis was at an angle $\theta = 70^\circ$ to the line of sight; all observed properties of B1 available at that time were given accurately by the model. The prediction of this model would be ${\cal R^\prime} = 1/0.42 = 2.4$, a factor of 5 larger than our result (table 2) for B1. Moreover, now that the \citet{TC08} survey of dark cloud cores has shown that B1 has the greatest $B_{LOS}$ of any core with a detected $B_{LOS}$, it seems more likely that {\bf B} is nearly along the line of sight, and that the true central total $B$ is close to the observed $B_{LOS} = 27$ $\mu$G and not the model result $B = 85$ $\mu$G, which implied $B_{LOS} = 85 cos 70^\circ = 29$ $\mu$G. When B1 was among a very small number of dark clouds with sensitive OH Zeeman observations, it was not unreasonable to hypothesize that its field lay nearly in the plane of the sky. However, other clouds similar to B1 with similar total field strengths should have {\bf B} nearly along the line of sight, yielding $B_{LOS} \sim 85$ $\mu$G; these are not found in the \citet{TC08} survey results. 

\citet{CB00} computed a model for L1544. The model had 30 $M_\odot$ within a radius of 0.45 pc, with additional mass that does not participate in the evolution in the envelope at larger radius (not clearly specified, but apparently about 2.5 pc based on their figure 1c). They assumed $\theta \approx 74^\circ$, again a very large angle between {\bf B} and the line of sight, which was necessary in order to have the required large central $B$ agree with the small observed value of $B_{LOS}$. The initial $M/\Phi$ was 0.8 of critical, or closer to critical than other ambipolar diffusion models discussed above. This would imply ${\cal R^\prime} \approx 1.25$, which differs from our measurement ${\cal R^\prime} = 0.46 \pm 0.43$, although not by a highly significant amount. However, as for B1, the large value for $\theta$ they had to assume in order to make the field strength of the model agree with the observation of $B_{LOS}$ seems unreasonably large. Not all {\bf B} can lie near the plane of the sky!

The ambipolar diffusion model results are all ${\cal R^\prime} > 1$, with the actual value depending on the initial assumed $M/\Phi$. But even if all clouds start only very slightly subcritical, which would in itself minimize the importance of ambipolar diffusion in cloud evolution, our results are not consistent with the ambipolar diffusion requirement.

A possible way out of this conclusion might be to hypothesize that the magnetic fields in our envelope regions are not the regular ones of the idealized models, but rather that these magnetic fields twist and indeed reverse direction. By fitting for a single value of $B_{LOS}$ for each envelope, we obtain a smaller value for $B_{LOS}$ (and hence a smaller ${\cal R^\prime}$) in the entire envelope surrounding the core than we would have gotten if we had (for example) just taken the largest value of $B_{LOS}$ at any of the four positions. If the field at the GBT positions twisted significantly, then perhaps the largest absolute value of $B_{LOS}$ would be the appropriate one to use in calculating the magnetic flux in the envelope region. At the other three positions surrounding a core, $B_{LOS}$ could be smaller due to {\bf B} being twisted into the plane of the sky or even to pointing in the opposite direction from the field in the core. However, our fit for a single envelope value of $B_{LOS}$ is the proper one for testing at least the idealized models of this theory. This single value gives (approximately) the result that would be obtained with a ring telescope beam as it would sample the envelope regions. But more importantly, the GBT beams are centered only $6^\prime$ from the center of each core, or at a radius of 0.26 pc for a typical cloud distance of 150 pc. This radius is about an order of magnitude smaller than the cloud radii in the three idealized ambipolar diffusion models discussed above, where the models require a smooth and not tangled magnetic field. The GBT positions are not far enough away from the cores that they could be sampling unrelated flux tubes that may have nothing to do with the envelopes of the cores. Moreover, such a twisted (indeed reversed) field explanation would have to hold for all four of our cores, which seems statistically unlikely. Finally, the good agreement in radial velocities of the core and envelope OH spectral lines (figures 2-5) suggests that the GBT observations do sample the envelope regions of the cores. 

What about predictions of initially weak magnetic field, turbulent theories? Such simulations are of course also idealized, with artificial boundary conditions, simple turbulence algorithms without a physical mechanism for generating the turbulence, and neglect of some physics. Nonetheless, our comparison is with state-of-the-art simulations. Recently, \citet{Letal08} have computed synthetic Zeeman profiles for their super-Alfv\'{e}nic (weak magnetic field) simulations of molecular clouds and cores formed by turbulence in an initially uniform interstellar medium. The mean field strength in their simulations is quite low;  $\overline{B}_{LOS} \approx 2.1$ $\mu$G along the direction of the original uniform {\bf B}. They performed numerically the same experiment for which we report here the observational results. That is, they ``observed'' the synthetic Stokes I and V spectra at each of their core positions with a $3^\prime$ beam and at four envelope positions around each core with $8^\prime$ beams pointed $6^\prime$ north, south, east, and west of the core positions. They computed $N(OH)$ from their Stokes I profiles and  $B_{LOS}$ by fitting their synthetic Stokes $dI/d\nu$ to their $V$ profiles (the same technique used observationally). They analyzed 36 (out of 139 total) cores with $B_{LOS} > 10$ $\mu$G, corresponding roughly to field strengths in the four cores we have observed. For these 36 cores, the mean $N(H_2) = 6.6 \times 10^{21}$ cm$^{-2}$, which agrees reasonably well with the mean  $N(H_2) = 4.3 \times 10^{21}$ cm$^{-2}$ for our four cores.  The mean line width of the 36 cores was $\Delta V \approx 1$ km s$^{-1}$, which also agrees with our four observed cores. For these cores they found that ${\cal R}$ had a large scatter, $0.08 \leq {\cal R} \leq 1.6$, with the PDF favoring values less than 1. Hence, the weak field, turbulent calculation of the formation of molecular clouds and cores seems to agree in physical properties with the four cores we have observed. Given that our experiment included only four clouds, the range in ${\cal R}$ corresponds well with the observed range. 

We suggest that other simulations, such as those with stronger magnetic fields that include both ambipolar diffusion and turbulence, should be compared with our experimental results in the same manner as did \citet{Letal08}. 

\section{Conclusion}

Previous Zeeman studies of magnetic fields in molecular clouds have not been definitive in testing the two extreme-case models of star formation. The mean mass-to-flux ratios $M/\Phi$ found from these statistical studies were slightly supercritical -- consistent with either theory. Detailed ambipolar diffusion models for two clouds found excellent agreement with the observations, although both required the field to be nearly in the plane of the sky in order not to produce line-of-sight fields much stronger than observed. Uncertainties in the angle between {\bf B} and the line of sight and in the total hydrogen column density are inherent in measuring  with Zeeman observations. In order to mitigate these uncertainties, we have measured the ratio of $M/\Phi$ between the envelopes and cores of four molecular clouds in order to test ambipolar diffusion (strong magnetic fields) versus turbulence (weak magnetic fields) driven star formation theory. The theory of star formation that hypothesizes clouds initially supported by strong magnetic fields, with evolution and core formation being driven by ambipolar diffusion, predicts that the central $M/\Phi$ must increase as ambipolar diffusion acts. Idealized models predict that the increase in $M/\Phi$ up to the point when the core becomes supercritical and gravitational collapse proceeds is approximately equal to the inverse of the amount by which the original cloud was subcritical; that is, ${\cal R^\prime} > 1$. The probability that all four of our clouds have ${\cal R^\prime} > 1$ is $3 \times 10^{-7}$, a highly significant result. On the other hand, simulations which form clouds and cores by turbulence acting in a weak magnetic field environment preferentially yield a $M/\Phi$ ratio between core and envelope ${\cal R} < 1$, in agreement with our results. 

Telescope availability limitations allowed only four clouds to be observed; unfortunately the extremely large amount of telescope time required precludes expanding this experiment beyond four clouds for the foreseeable future. The theoretical predictions of ${\cal R}$ and ${\cal R^\prime}$ are based on idealized ambipolar diffusion models and idealized turbulence simulations. Nonetheless, the clear conclusion from our experiment is that at least for these four clouds, the prediction of the idealized ambipolar diffusion models does not agree with our observational results, while the prediction of initially supercritical turbulence-driven simulations does. Still untested is whether simulations that include both significant magnetic fields and turbulence better match the data than either of the extreme cases. We suggest that all theorists who simulate the formation and evolution of molecular clouds and cores test their simulations against the results of this experiment in the manner of \citet{Letal08}; that is, by calculating Stokes I and V spectra of OH from the simulations and ``observing'' $B_{LOS}$ with our beam patterns (figure 1). 

\acknowledgments
We thank the National Radio Astronomy Observatory for the allocation of about 300 hours of observing time on the GBT in response to proposal GBT07A-029, and the observatory staff for help in making the observations successful. We thank Phil Perillat of the Arecibo Observatory for help in re-calibrating the Arecibo spectra, and we especially thank Carl Heiles and Tim Robishaw for allowing us to use their GBT software and for helping us understand how to make it work. Finally, we thank two conscientious referees who provided valuable suggestions to refine the scientific and statistical arguments in this paper. This research was partially supported by NSF grants AST 0307642 and 0606822, and by NRAO grant GSSP07-0007.

\facility{{\it Facilities} Arecibo, GBT}



\clearpage

\begin{deluxetable}{lccc}
\tablecaption{Observational Results}
\tablewidth{0pt}
\tablehead{
\colhead{Cloud} & \colhead{$I_{OH} (K)$}& \colhead{$\Delta V (km/s)$}& \colhead{$B_{LOS} (\mu G)$}
}
\startdata
L1448CO(env)   &  0.63 & 1.16  & $-0\pm5$ \\
L1448CO(core)  &  1.30 & 0.93  & $-26\pm4$ \\
\\
B217-2(env)  &  0.59 & 0.79  & $+2\pm4$ \\
B217-2(core) &  1.28 & 0.47  & $+14\pm4$ \\
\\
L1544(env)   &  0.96  & 0.67  & $+2\pm3$ \\
L1544(core)  &  2.43  & 0.48  & $+11\pm2$ \\
\\
B1(env)         &  1.21  & 1.32  & $-8\pm3$ \\
B1(core)       &   1.93  &  1.14 & $-27\pm4$ \\
\enddata
\end{deluxetable}

\begin{deluxetable}{lccc}
\tablecaption{Relative Mass/Flux}
\tablewidth{0pt}
\tablehead{
\colhead{Cloud} &  \colhead{${\cal R}$} & \colhead{${\cal R^\prime}$} & \colhead{Probability ${\cal R}$ or ${\cal R^\prime} > 1$} }
\startdata
L1448CO   &  $0.02 \pm 0.36$  & $0.07 \pm 0.34$ & 0.005 \\
B217-2  &  $0.15 \pm 0.43$ & $0.19 \pm 0.41$ & 0.05 \\
L1544   &  $0.42 \pm 0.46$ & $0.46 \pm 0.43$ & 0.11\\
B1         & $ 0.41 \pm 0.20$  & $0.44 \pm 0.19$ & 0.010\\
\enddata
\end{deluxetable}

\clearpage
\begin{figure}
\epsscale{0.5}
\plotone{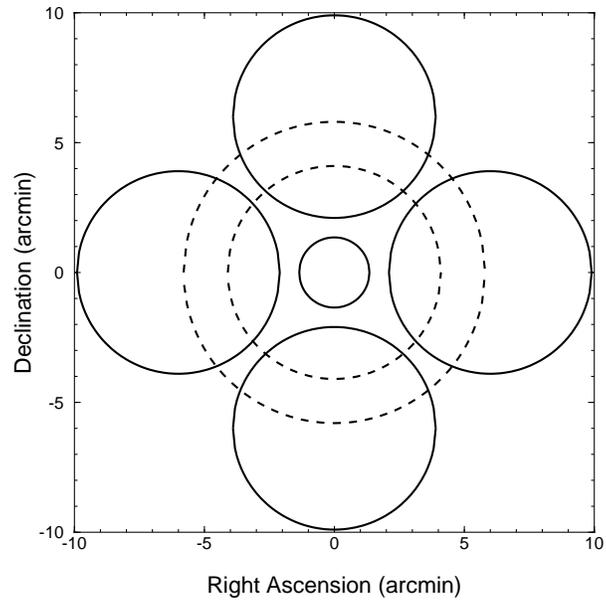}
\caption{The Arecibo telescope primary beam (small circle centered at 0,0) and the four GBT telescope primary beams (large circles centered $6^\prime$ north, south, east, and west of 0,0. The dotted circles show the first sidelobe of the Arecibo telescope beam. All circles are at the half-power points.}
\label{f1}
\end{figure}

\clearpage
\begin{figure}
\epsscale{1.0}
\plotone{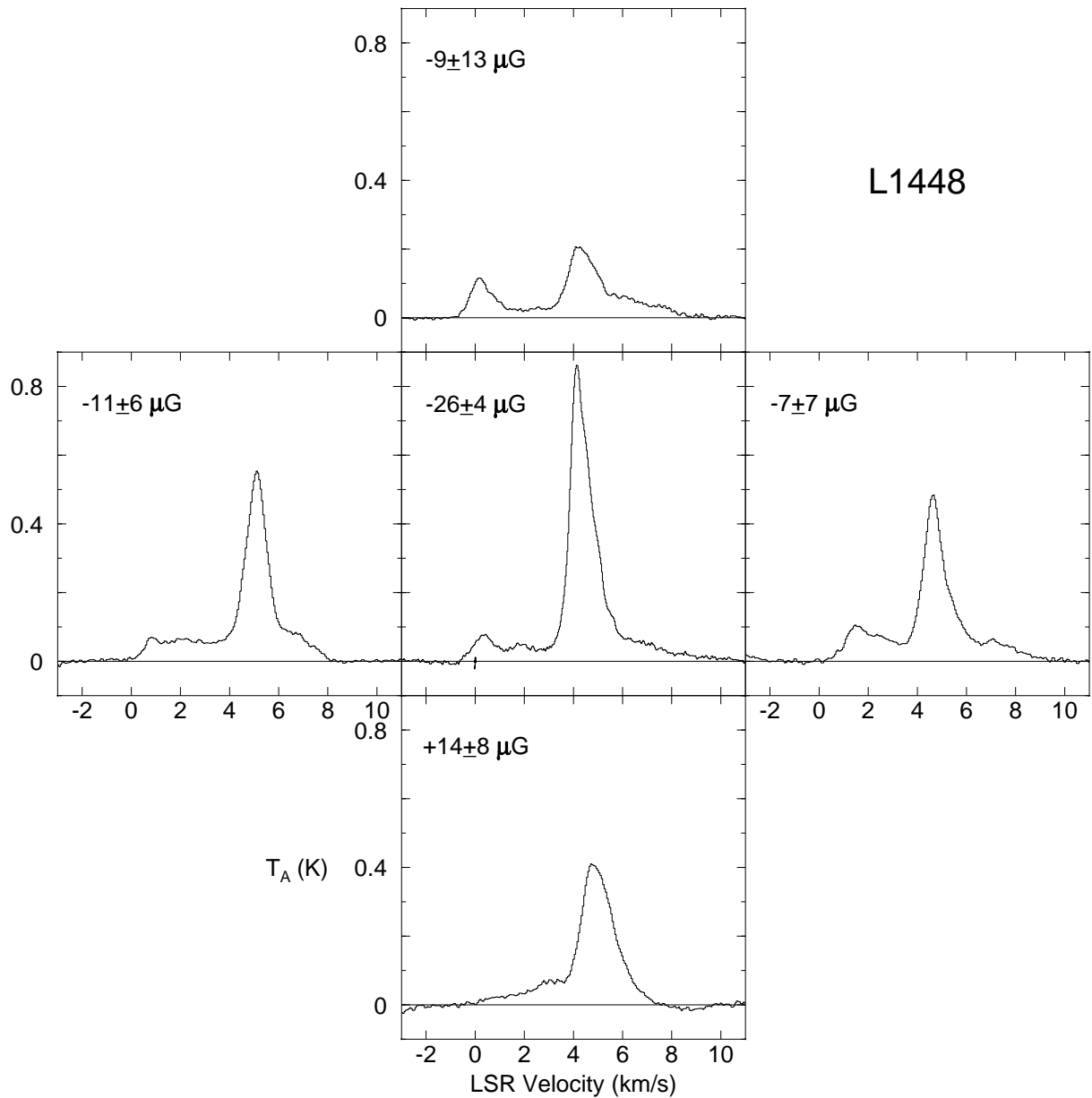}
\caption{OH 1667 MHz spectra toward the core of L1448CO obtained with the Arecibo telescope (center panel) and toward each of the envelope positions $6^\prime$ north, south, east, and west of the core, obtained with the GBT. In the upper left of each panel is the inferred $B_{LOS}$ and its $1\sigma$ uncertainty at that position. A negative $B_{LOS}$ means the magnetic field points toward the observer, and vice versa for a positive $B_{LOS}$.}
\label{f2}
\end{figure}

\clearpage
\begin{figure}
\epsscale{1.0}
\plotone{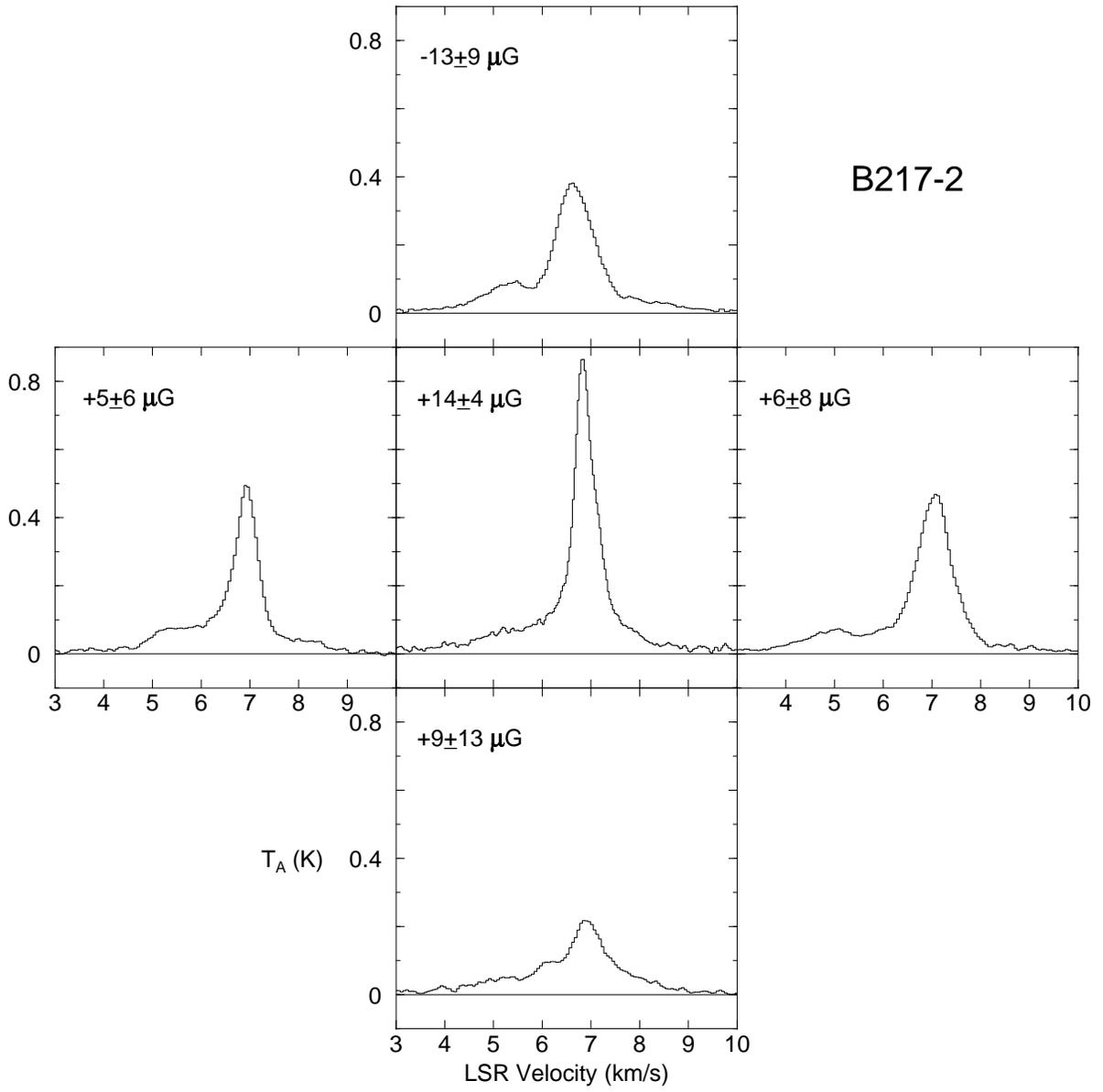}
\caption{As in figure 2, but for B217-2.}
\label{f3}
\end{figure}

\clearpage
\begin{figure}
\epsscale{1.0}
\plotone{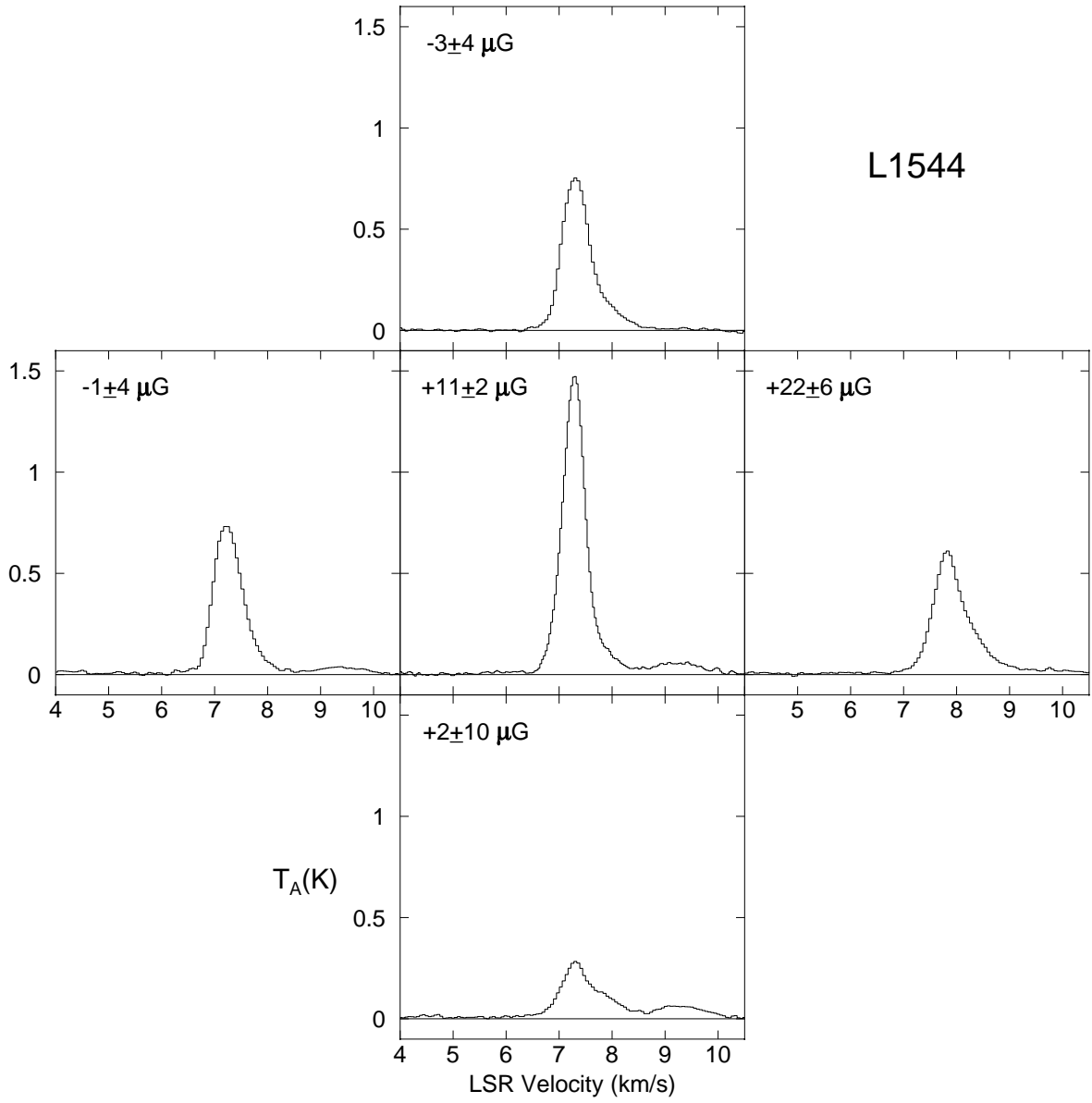}
\caption{As in figure 2, but for L1544.}
\label{f4}
\end{figure}

\clearpage
\begin{figure}
\epsscale{1.0}
\plotone{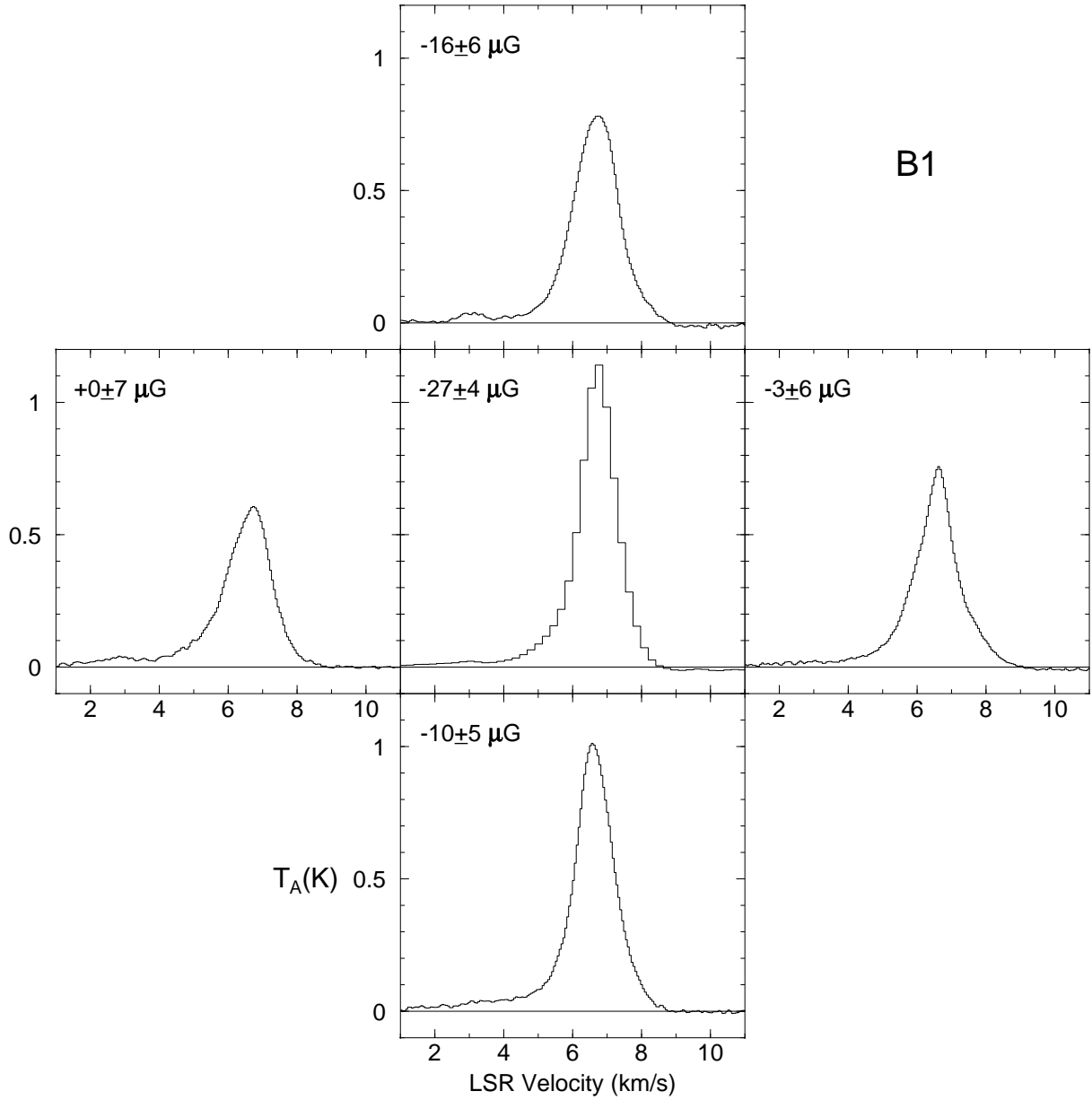}
\caption{As in figure 2, but for B1.}
\label{f5}
\end{figure}

\clearpage
\begin{figure}
\centering
\includegraphics[width=.4\textwidth]{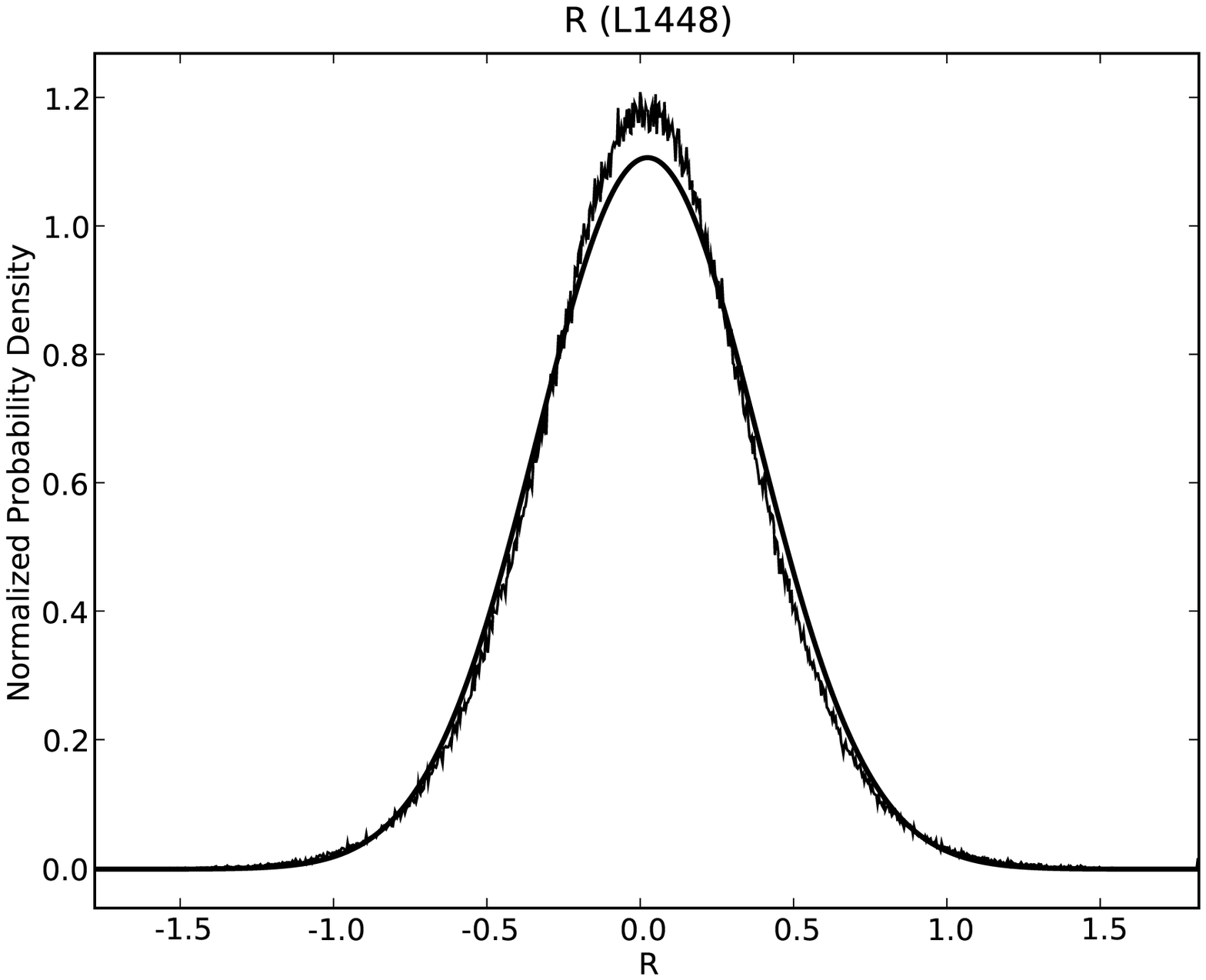}\hspace*{10mm}
\includegraphics[width=.4\textwidth]{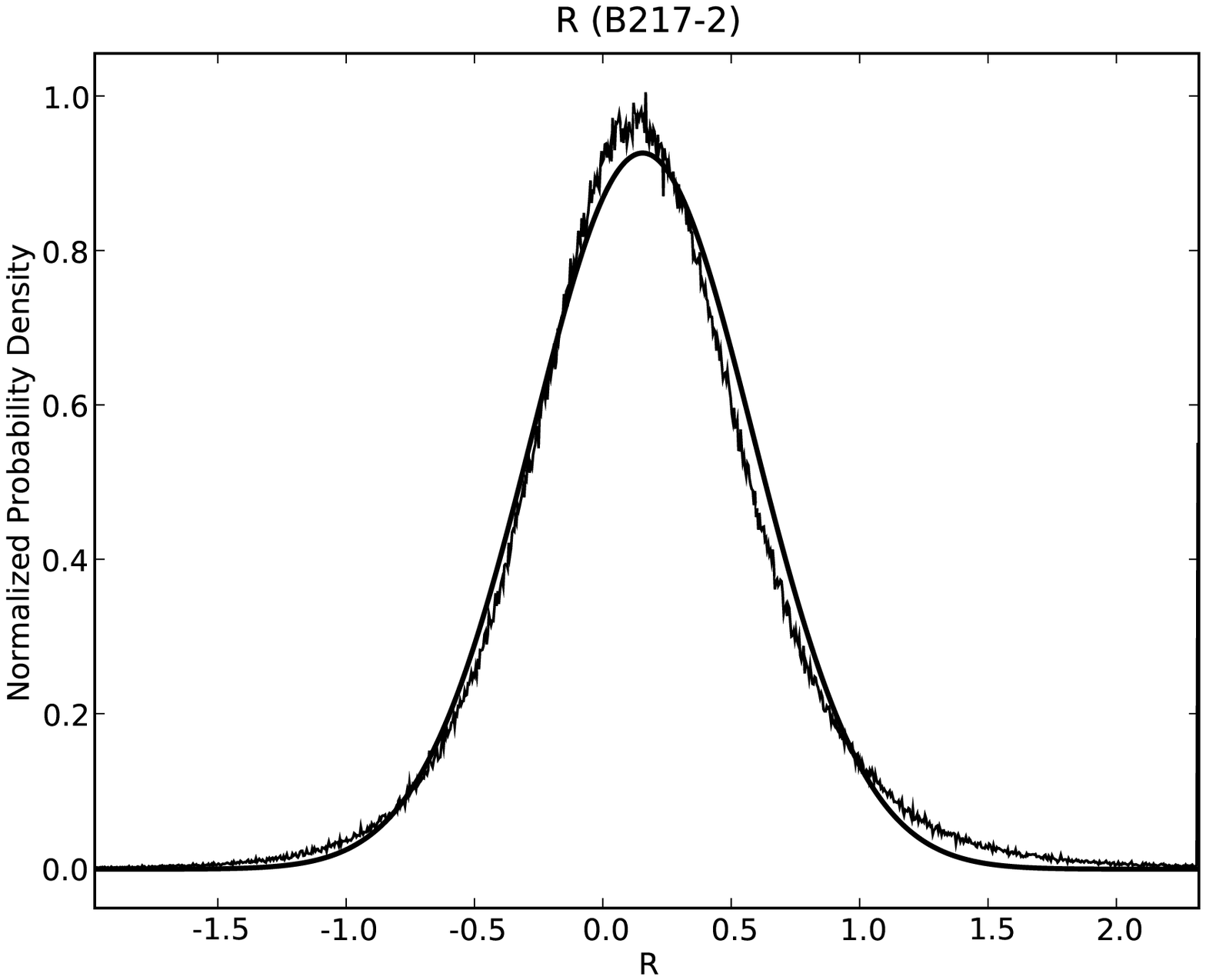}\\[5mm]
\includegraphics[width=.4\textwidth]{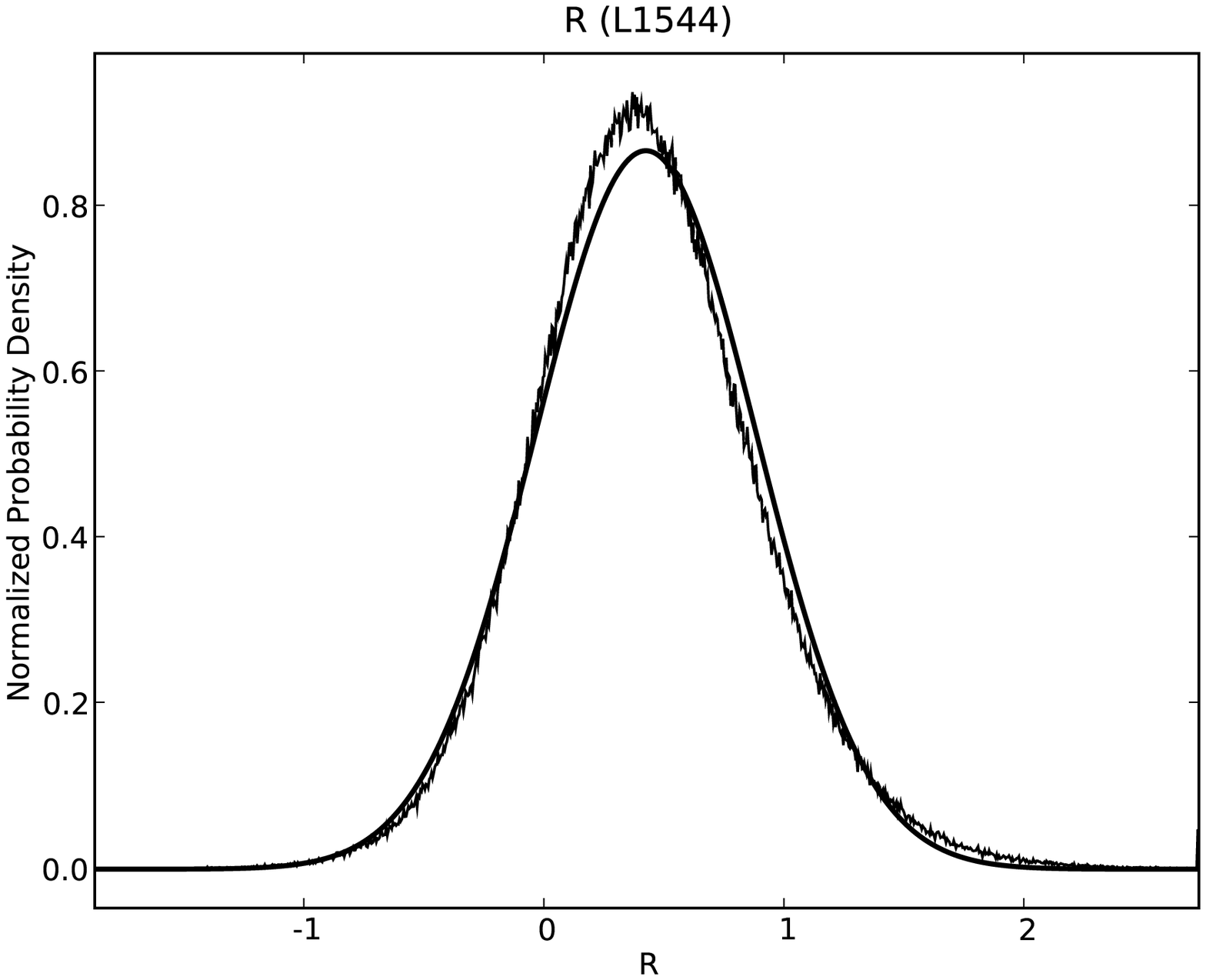}\hspace*{10mm}
\includegraphics[width=.4\textwidth]{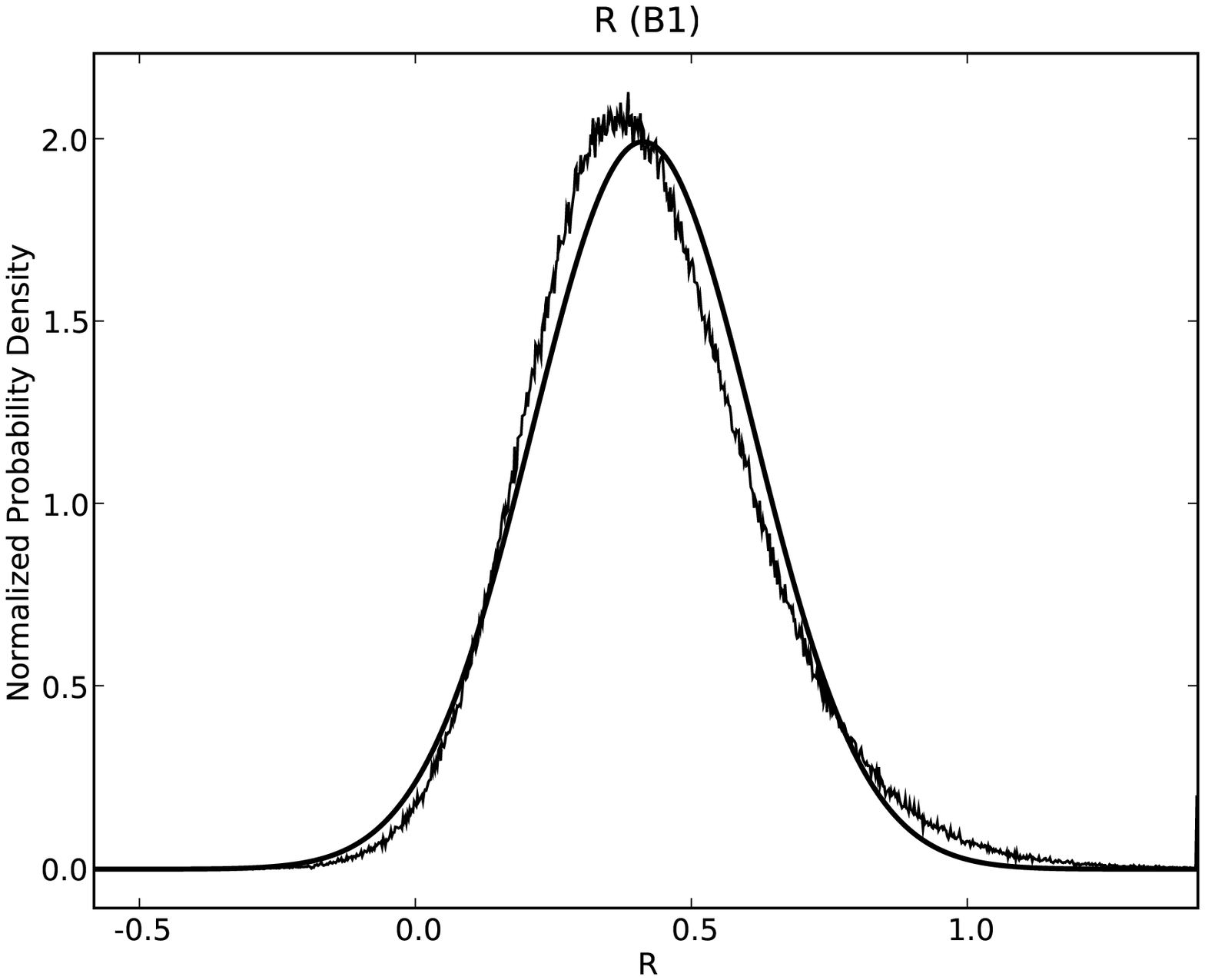}
\caption{Monte Carlo estimate of ${\cal R}$ for each cloud. The Gaussian normal error curve for the error propagation results given in table 2 are also shown.}
\label{f6}
\end{figure}

\end{document}